\documentclass[aps,prb,showpacs,twocolumn,groupedaddress]{revtex4}

\usepackage[T1]{fontenc}
\usepackage[latin1]{inputenc}
\usepackage{latexsym}
\usepackage{array}
\usepackage{amsfonts}
\usepackage[centertags]{amsmath}
\usepackage{bm}
\usepackage{wasysym}
\usepackage[dvips,xdvi]{graphicx}
\usepackage{textcomp}
\usepackage{epsf}
\usepackage{hhline}

\begin{document}

\title{Nature of the induced ferroelectric phases in $\textrm{TbMn}_2\textrm{O}_5$}

\author{P.~Tolédano}\affiliation{Laboratory of Complex Systems, 33 rue Saint-Leu, University of Picardie, 80000 Amiens, France}

\author{W.~Schranz}\affiliation{Faculty of Physics, University of Vienna,
Boltzmanngasse\,5, A-1090 Vienna, Austria}

\author{G.~Krexner}\affiliation{Faculty of Physics, University of Vienna,
Boltzmanngasse\,5, A-1090 Vienna, Austria}

\date{\today}

\begin{abstract}
\noindent
The magnetostructural transitions and magnetoelectric effects reported in $\textrm{TbMn}_2\textrm{O}_5$ are described theoretically and shown to correspond to two essentially different mechanisms for the induced ferroelectricity. The incommensurate and commensurate phases observed between 38 K and 24 K exhibit a hybrid pseudo-proper ferroelectric nature resulting from an effective bilinear coupling of the polarization with the antiferromagnetic order-parameter. This explains the high sensitivity of the dielectric properties of the material under applied magnetic field.  Below 24 K the incommensurate phase shows a standard improper ferroelectric character induced by the coupling of two distinct magnetic order-parameters. The complex dielectric behavior observed in the material reflects the crossover from one to the other transition regime. The temperature dependences of the pertinent physical quantities are worked out and previous theoretical models are discussed.

\pacs{61.50.Ah, 77.80.-e, 75.80.+q}

\end{abstract}

\maketitle

\section{Introduction}

It was recently observed\cite{KimuraNature,KimuraMatter} that an electric polarization can emerge at a magnetic transition if the magnetic spins order in non-collinear spiral structures. This new type of magnetostructural transition was reported in various classes of  multiferroic materials\cite{Fiebig,Eerenstein,Ramesh,Cheong}, such as the rare-earth manganites $\textrm{RMnO}_3$\cite{Goto} and $\textrm{RMn}_2\textrm{O}_5$\cite{Hur1,Chapon1}, $\textrm{Ni}_3\textrm{V}_2\textrm{O}_8$\cite{Lawes}, $\textrm{MnWO}_4$\cite{Taniguchi}, $\textrm{CoCr}_2$\cite{Tomiyasu} or $\textrm{Cr}_2\textrm{BeO}_4$\cite{Newnham}. In these compounds the correlation between spins and electric dipoles gives rise to remarkable magnetoelectric effects, indicating a strong sensitivity to an applied magnetic field, such as reversals or flops of the polarization, and a strong enhancement of the dielectric permittivity. In the aforementioned materials the ferroelectric phases appear below an intermediate non-polar magnetic phase, i.e. the breaking of inversion symmetry, which allows emergence of ferroelectric properties, does not occur simultaneously with the breaking of time reversal symmetry.

Theoretical arguments have been raised\cite{Katsura,Mostovoy,Sergienko,Kenzelmann,JHu} to justify the observation of magnetoelectric effects in spiral magnets. However, a comprehensive theoretical description of the experimental results disclosed in multiferroic materials could not be achieved because the actual symmetries of the primary (magnetic) and secondary (structural) order-parameters have not been related organically to the thermodynamic functions which provide the relevant phase diagrams. Here, we give a unifying theoretical description of the magnetostructural transitions found in the manganite $\textrm{TbMn}_2\textrm{O}_5$\cite{Hur1,Chapon1,Blake,Koo} in the framework of the Landau theory of magnetic phase transitions\cite{Landau,Dzialo,Toledano1}. It reveals that the transitions observed in this compound at 38 K and 24 K correspond to essentially different mechanisms for the induced ferroelectricity: The 38 K transition involves an \textit{effective bilinear} coupling of the polarization with a \textit{single} magnetic order-parameter. It results in a \textit{pseudo-proper} ferroelectric nature for the phases stable between 38 K and 24 K. By contrast, the 24 K transition exhibits an \textit{improper} ferroelectric behavior corresponding to a linear-quadratic coupling of the polarization with \textit{two distinct} magnetic order-parameters.  The crossover from one to the other transition mechanism provides an interpretation of the  dielectric behavior observed in absence or presence of an applied magnetic field\cite{Hur1,Chapon1}.

\begin{table*}
\includegraphics[scale=0.9]{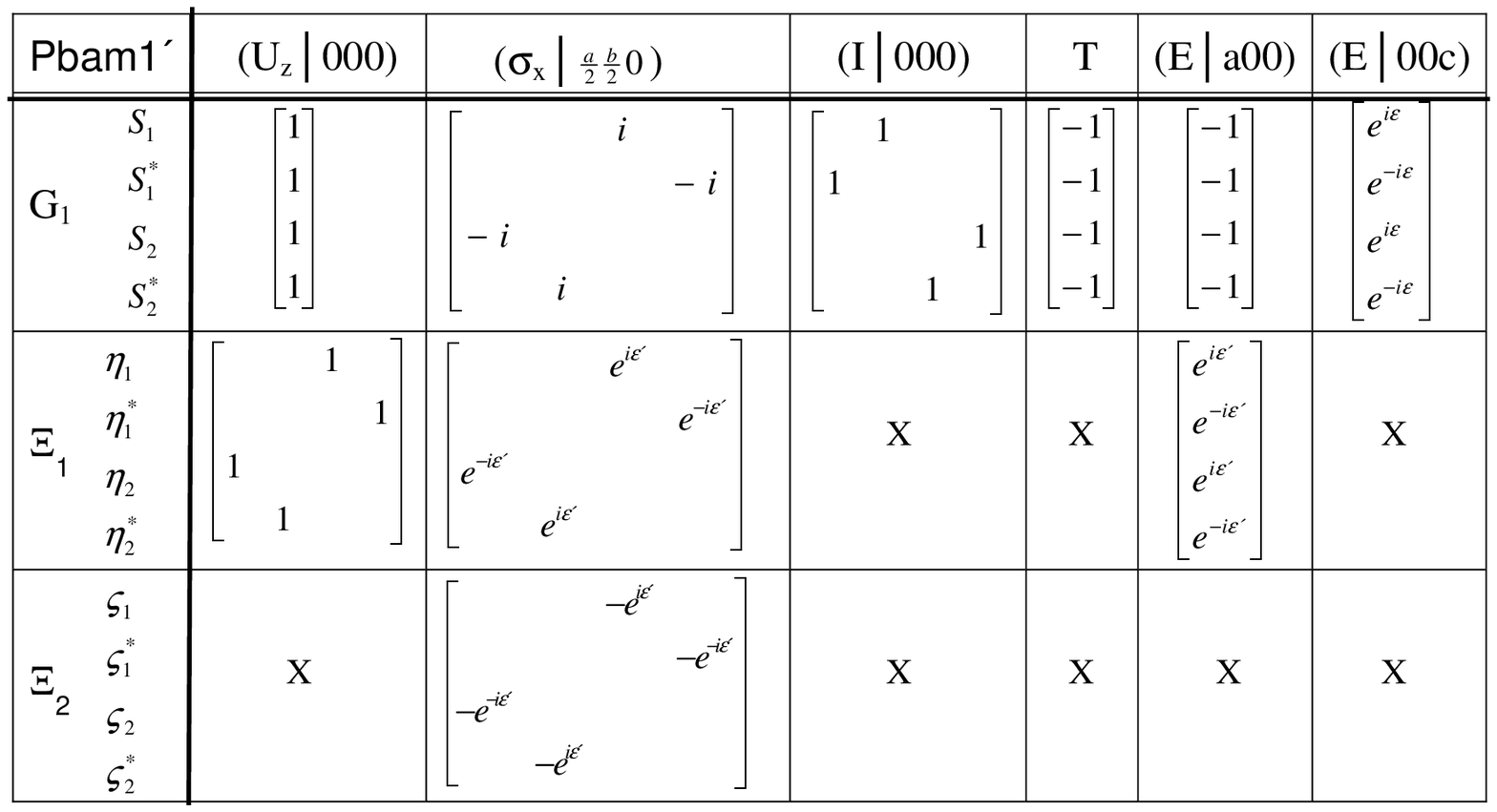}
\caption{Generators of the IC's $G_1$, $\Xi_1$ and $\Xi_2$. Diagonal $4\times4$ matrices are represented by columns. A cross ($\times$) indicates that the matrix is the same as in the upper row. $\epsilon=k_zc$ except in the commensurate phase III where $\epsilon=\frac{\pi}{2}$. $\epsilon'=k_x\frac{a}{2}$  . $G_1$ is deduced from the irreducible representation (IR) of the group $G_k$=mm2, denoted $\hat{\tau}_1(k_{16})$ in Kovalev's tables\cite{Kovalev}. $\Xi_1$ and $\Xi_2$ are deduced from the IR's ($\hat{\tau}_1(k_3)$ and $\hat{\tau}_2(k_3)$) of $G_k=m_y$.}
\label{fig:tab}
\end{table*}

 On cooling below the paramagnetic Pbam1' (P) structure, $\textrm{TbMn}_2\textrm{O}_5$ undergoes five phase transitions\cite{Hur1,Chapon1} taking place successively at $T_1=43$ K, $T_2=38$ K, $T_3=33$ K, $T_4=24$ K and $T_5=10$ K, the corresponding phases being denoted I to V.
 The paper is organized as follows: In Sec. II the P$\rightarrow$ I $\rightarrow$ II $\rightarrow$ III sequence of transitions giving rise at $T_1$, $T_2$ and $T_3$, to the incommensurate phases I and II, and commensurate phase III\cite{Hur1,Chapon1}, is described theoretically. The remarkable magnetoelectric effects ocurring at the III$\rightarrow$ IV $\rightarrow$ V transitions, taking place at $T_4$ and $T_5$, are analyzed in Sec. III. In Sec. IV our results are summarized and compared to the results obtained in previous theoretical works on $TbMn_2O_5$ \cite{Radaelli2007,Radaelli2008,Harris2007,Harris2008a,Harris2008b,Kadomtseva,Sushkov}. Adaption of our description to other $RMn_2O_5$ compounds \cite{Noda,Chapon2,Bodenthin,Higashiyama,Hur2004,Munoz} is outlined.

 \section{The P$\rightarrow$ I $\rightarrow$ II $\rightarrow$ III Transitions}

 The wave-vector associated with the incommensurate antiferromagnetic phase I and ferroelectric phase II is $\vec{k}=(1/2,0,k_z)$, with $k_z$ decreasing from about 0.30 to 0.25. It is associated with a 4-dimensional irreducible corepresentation (IC) of Pbam1', denoted $G_1$, whose generators are given in Table I. The complex amplitudes $S_1=\rho_1e^{i\theta_1}$, $S_1^*=\rho_1e^{-i\theta_1}$, $S_2=\rho_2e^{i\theta_2}$, $S_2^*=\rho_2e^{-i\theta_2}$ of the magnetic waves transforming according to $G_1$, form the symmetry-breaking order-parameter for the P$\rightarrow$ I $\rightarrow$ II transitions, giving rise to the invariants $\mathfrak{I}_1=\rho_1^2+\rho_2^2$, $\mathfrak{I}_2=\rho_1^2\rho_2^2$, and $\mathfrak{I}_3=\rho_1^2\rho_2^2\textrm{cos}2\theta$, with $\theta=\theta_1-\theta_2$. Therefore the homogeneous part of the free-energy density reads:
\begin{eqnarray}
&&\Phi_1(\rho_1,\rho_2,\theta) = \\ \nonumber
&&a_1\mathfrak{I}_1+a_2\mathfrak{I}_1^2+b_1\mathfrak{I}_2+
b_2\mathfrak{I}_2^2+c_1\mathfrak{I}_3+c_2\mathfrak{I}_3^2+d\mathfrak{I}_1\mathfrak{I}_3+...
\label{freeenergydensity}
\end{eqnarray}

An eighth degree expansion is required in order to account for the full set of stable phases
resulting from the minimization of $\Phi_1$ and for disclosing the magnetoelectric properties observed in $TbMn_2O_5$. It stems from the following rule demonstrated in Ref. \onlinecite{Toledano2}: If n is the highest degree of the basic order-parameter invariants (here n=4 for the $\mathfrak{I}_2$ and $\mathfrak{I}_3$ invariants), the free energy has to be truncated at not less than the degree 2n (here 2n=8) for ensuring the stability of all phases involved in the phase diagram. However, one can neglect most of the non-independent invariants of degrees lower or equal to eight (as for example $\mathfrak{I}_1^3$, $\mathfrak{I}_1^4$, $\mathfrak{I}_1\mathfrak{I}_2$ or $\mathfrak{I}_2\mathfrak{I}_3$) which can be shown to have no influence on the stability of the phases, but only modify secondary features of the phase diagram, as for example the shape of the transition lines separating the stable phases. In contrast the invariant
$\mathfrak{I}_1\mathfrak{I}_3$ has to be taken into account for stabilizing "asymmetric" phases with $\rho_1 \neq \rho_2$. Note that the fourth-degree invariants $\rho_1^2\rho_2^2$ and $\rho_1^2\rho_2^2cos2\theta$ express at a phenomenological level the exchange striction interactions and anisotropic exchange forces, respectively. Minimizing $\Phi_1$ with respect to $\theta$ yields the following equation of state:

\begin{equation}
\label{eq of state}
\rho_1^2\rho_2^2sin2\theta\left[c_1+d_1(\rho_1^2+\rho_2^2)+2c_2\rho_1^2\rho_2^2cos2\theta\right]=0
\end{equation}

Eq.(\ref{eq of state}) and the equations minimizing $\Phi_1$ with respect to $\rho_1$, $\rho_2$ show that \textit{seven} phases, labeled 1-7, can be stabilized below the P phase for different equilibrium values of $\rho_1$, $\rho_2$ and $\theta$.
Fig.~1 summarizes the equilibrium properties of each phase and their magnetic point-group symmetries. One can verify that the phases denoted 2,4,6 and 7 display a ferroelectric polarization along $y$ and that all phases correspond to an antiferromagnetic ordering except phases 5 and 7 which show a non-zero magnetization along $x$. The respective location of the phases is indicated in the theoretical phase diagrams shown in Fig.~2, in the space $(a_1,b_1,c_1)$ (Fig.~2(a)) and plane $(b_1,c_1)$ (Fig.~2(b)) of the phenomenological coefficients, and in the orbit space $(\mathfrak{I}_1,\mathfrak{I}_2,\mathfrak{I}_3)$ (Fig.~2(c)). It allows to determine the possible sequences of phases separated by second-order transitions as P$\rightarrow$1$\rightarrow$6$\rightarrow$7 or P$\rightarrow$3$\rightarrow$4$\rightarrow$7.
Dielectric and magnetic properties of the phases are deduced from the coupling of the order-parameter with the polarization $({\vec P})$- and magnetization $({\vec M})$- components, which are   $\mathfrak{I}_4=\rho_1\rho_2P_y\textrm{sin}\theta$, $\mathfrak{I}_5=(\rho_1^2-\rho_2^2)M_xM_y$ and $\mathfrak{I}_6=\rho_1\rho_2M_xM_z\textrm{cos}\theta$.
\begin{figure}
\begin{center}
\includegraphics[scale=0.8]{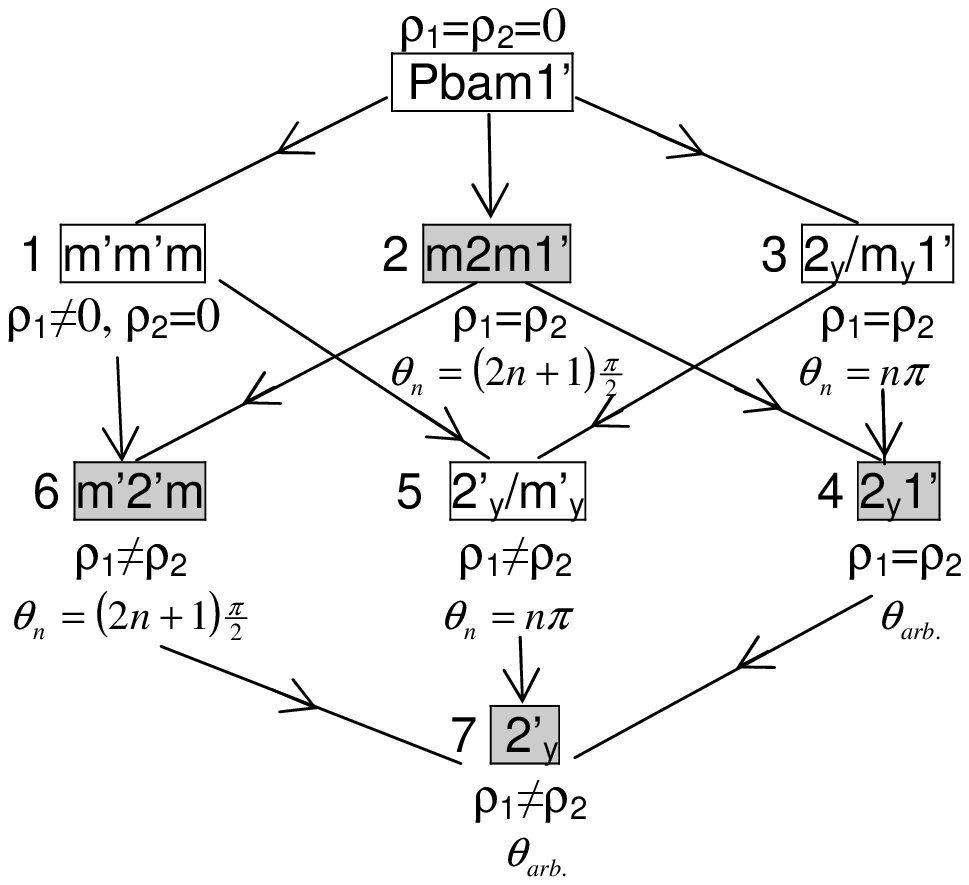}
\caption{Connections between the magnetic point-groups of phases 1-7 induced by the IC $G_1$ of Pbam 1', and equilibrium conditions fulfilled by the order-parameter in each phase. Gray rectangles indicate ferroelectric phases. $\theta_{arb.}$ stands for arbitrary.}
\label{fig:1}
\end{center}
\end{figure}

\begin{figure}
\begin{center}
\includegraphics[scale=0.4]{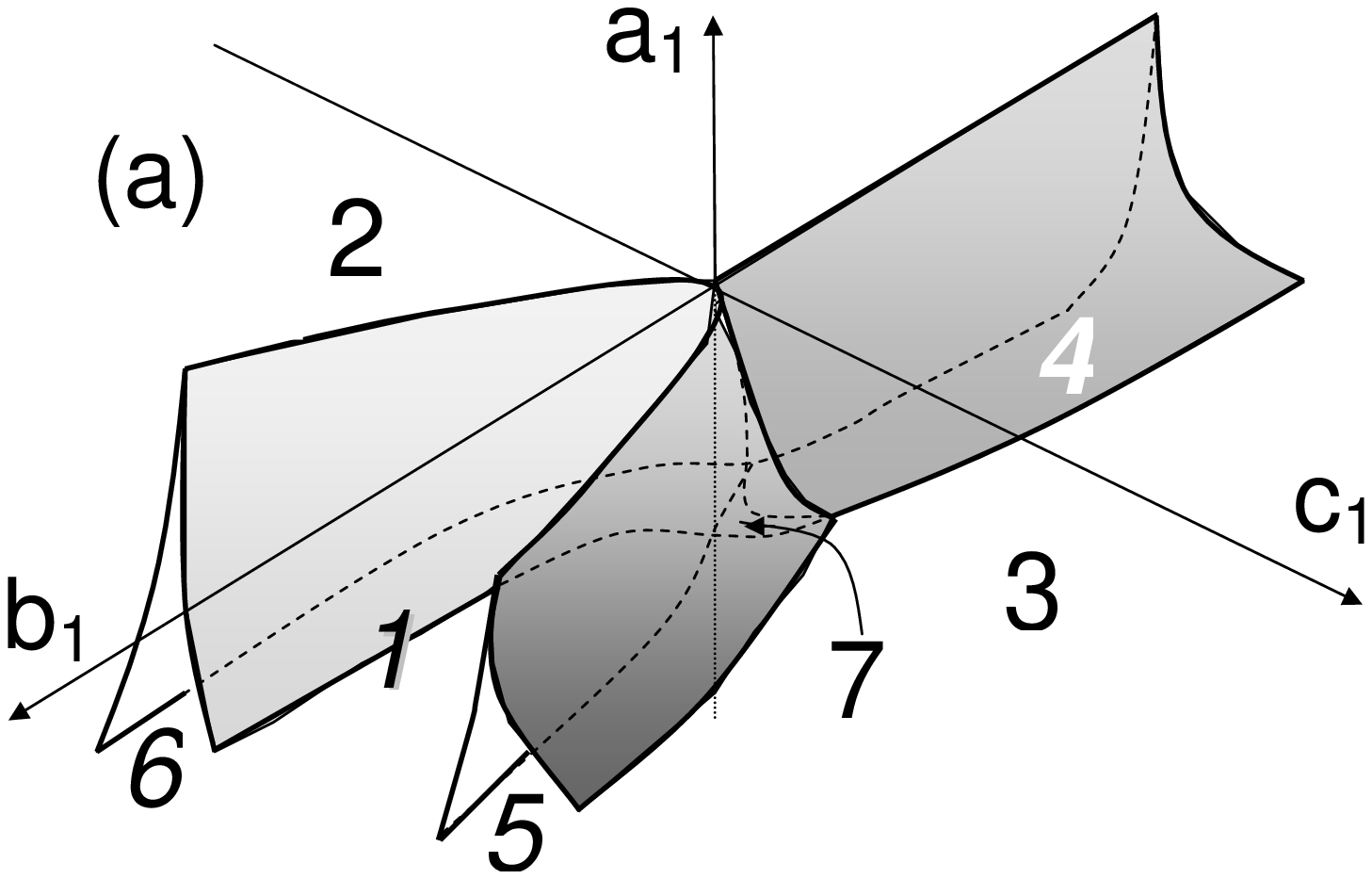}
\includegraphics[scale=0.35]{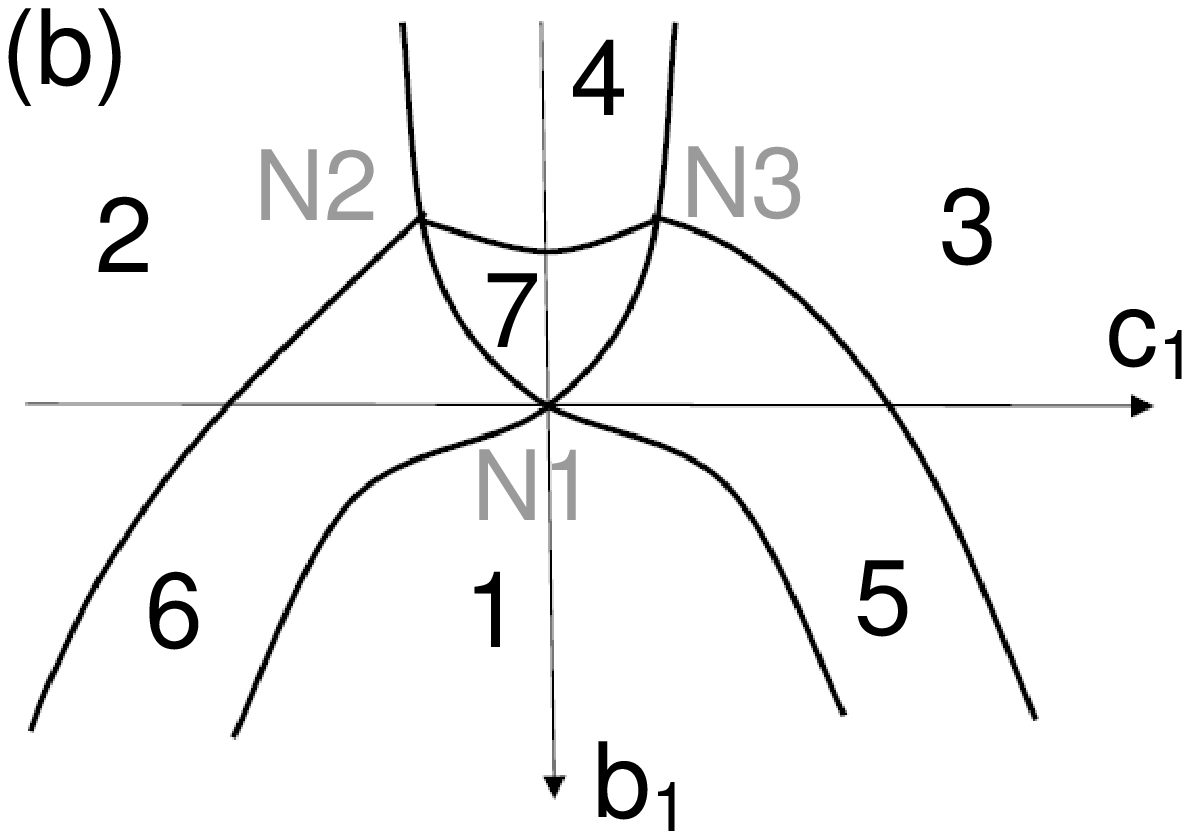}
\includegraphics[scale=0.30]{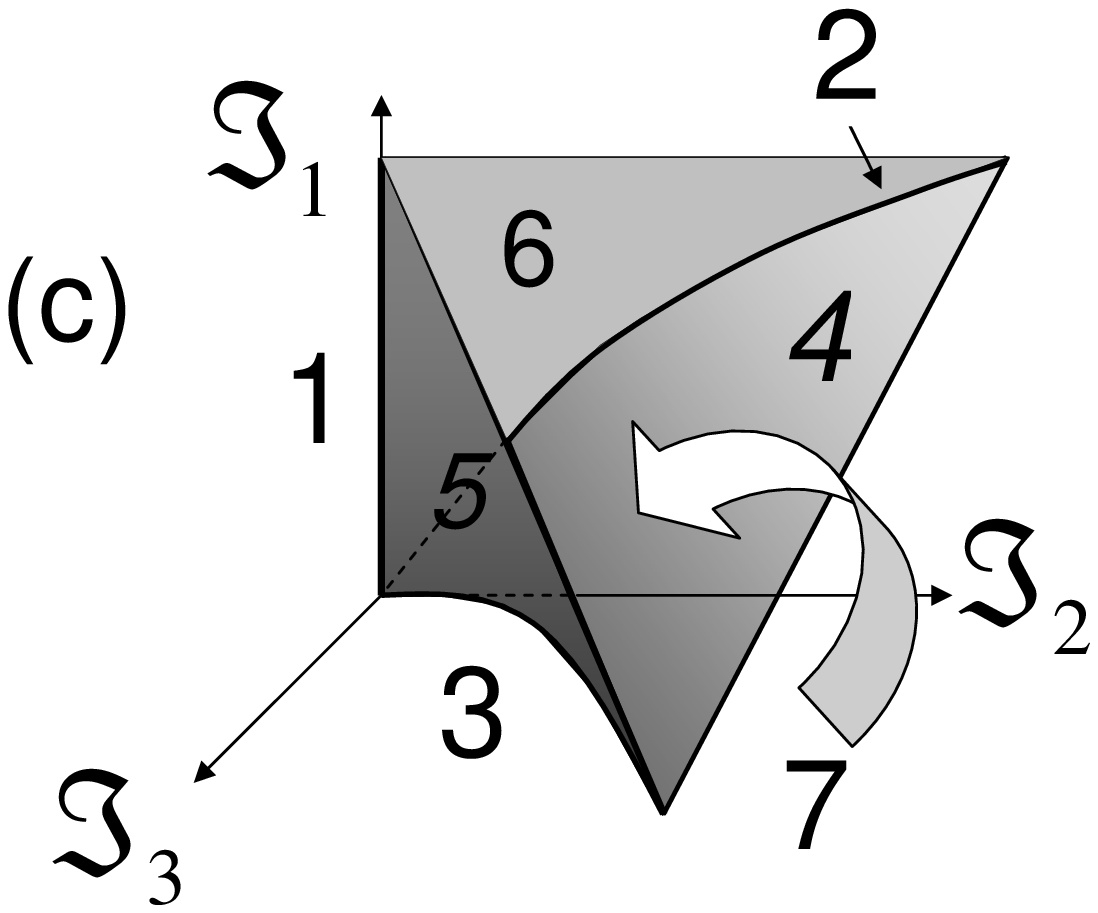}
\caption{Phase diagrams deduced from the minimization of the free-energy $\Phi_1$ given by Eq.~(1) in: (a) the ($a_1$, $b_1$, $c_1$) space, (b) the ($b_1$, $c_1$) plane for $a_1<0$ , and (c) the orbit space ($\mathfrak{I}_1$, $\mathfrak{I}_2$,$\mathfrak{I}_3$). In (a) the phases are separated by second-order transition surfaces, which become curves in (b). Phases 1, 2 and 3 can be reached directly from the P-phase across the second-order plane $a_1= 0$. In (b) N1, N2 and N3 are four-phase points, which become curves in (a). In (c) phases (1, 2, 3) and (4, 5, 6) correspond, respectively, to curves and surfaces. Phase 7 coincides with the volume limited by the surfaces (4, 5, 6).}
\label{fig:2}
\end{center}
\end{figure}
The preceding results allow a consistent interpretation of the P$\rightarrow$ I $\rightarrow$ II sequence of phases reported in $\textrm{TbMn}_2\textrm{O}_5$. Phase I observed between $T_1$ and $T_2$, corresponds to phase 1 ($\rho_1 \neq 0$, $\rho_2 = 0$) in Fig.~1. It displays the m'm'm symmetry with antiferromagnetic order in the (x,y) plane ($\mathfrak{I}_5=\rho_1^2M_xM_y$), a doubling of the lattice parameter $a$ and an incommensurate modulation along $c$, expressed by the Lifshitz invariant $\rho_1^2 \frac{\partial \theta_1}{\partial z}$ . Figs.~1 and 2 show that a continuous transition can occur from phase 1 to the ferroelectric phase 6 ($\rho_1 \neq \rho_2$, $\theta=(2n+1)\frac{\pi}{2}$ ), which exhibits a spontaneous polarization along $y$, and a magnetic symmetry m'2'm preserving an antiferromagnetic order in the (x,y) plane ($\mathfrak{I}_5 \neq 0$). Identifying phase 6 with phase II of $\textrm{TbMn}_2\textrm{O}_5$ allows a straightforward interpretation of the dielectric behavior observed at the I $\rightarrow$ II transition. From the dielectric free-energy density $\Phi_1^D=\delta_1 \rho_1\rho_2P_y\textrm{sin}\theta+\frac{P_y^2}{2\epsilon^0_{yy}}$, one gets the equilibrium value of $P_y$ in phase 6
\begin{equation}
\label{polarization}
P^e_y=\pm\delta_1\epsilon^0_{yy}\rho_1\rho_2
\end{equation}
The $(S_1,S_1^*)$ components related to $\rho_1$ have been already activated in phase 1, and \textit{are frozen in phase 6} , which is induced by the sole symmetry breaking mechanism related to $\rho_2$. Therefore, Eq.~(\ref{polarization}) reflects an \textit{effective bilinear coupling} of $P_y$  with $\rho_2$, giving rise to a \textit{proper ferroelectric} critical behavior at the transition between phases 1 and 6. This situation is reminiscent of \textit{pseudo-proper ferroelectric transitions}\cite{Toledano1} where the spontaneous polarization has the same symmetry as the transition order-parameter, to which it couples bilinearly, but results from an induced mechanism. In Phase II of $\textrm{TbMn}_2\textrm{O}_5$, $P_y$ and $\rho_2$ are related by a \textit{pseudo-proper-like} coupling since they display different symmetries. Therefore, at the I $\rightarrow$ II transition, $P_y$ varies critically as $\rho_2$, i.e. $P_y\propto(T_2-T)^{1/2}$, whereas the dielectric permittivity $\epsilon_{yy}$ exhibits a Curie-Weiss-like divergence $\epsilon_{yy}\propto\mid T-T_2 \mid^{-1}$. Figs.~3(a) and 3(b)  show the excellent fit of the experimental curves reported by Hur et al.\cite{Hur1} with the preceding power-laws. The \textit{induced} character of $P_y$ appears only from its magnitude (40 nC $\textrm{cm}^{-2}$)\cite{Hur1,Chapon1}, which is two orders smaller than in proper ferroelectrics. Note that a conventional trilinear (improper) coupling between $P_y$ and the magnetic order-parameters $\rho_1$ and $\rho_2$ would lead to an \textit{upward step} of $\epsilon_{yy}$ and to a \textit{linear} dependence of $P_y \propto (T_2-T)$.

\begin{figure}
\begin{center}
\includegraphics[scale=0.4]{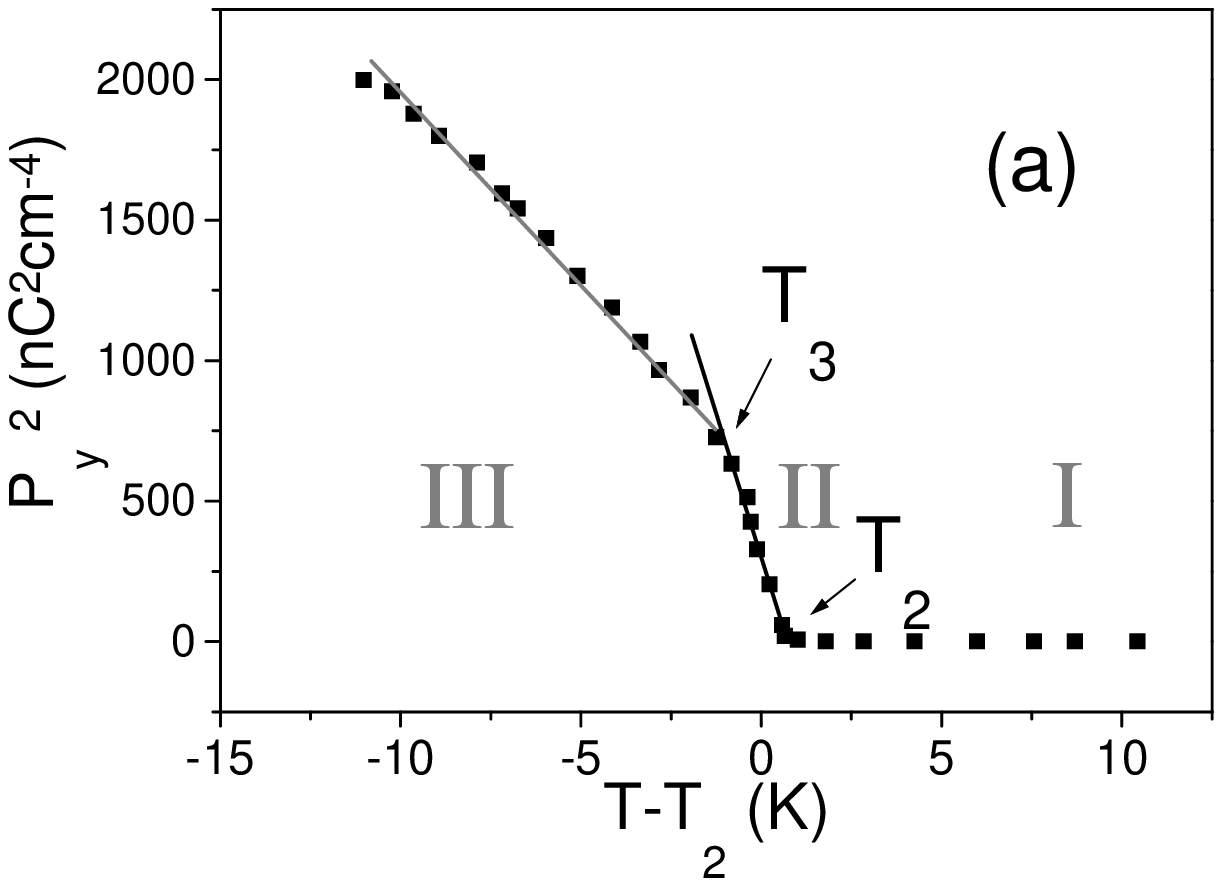}
\includegraphics[scale=0.4]{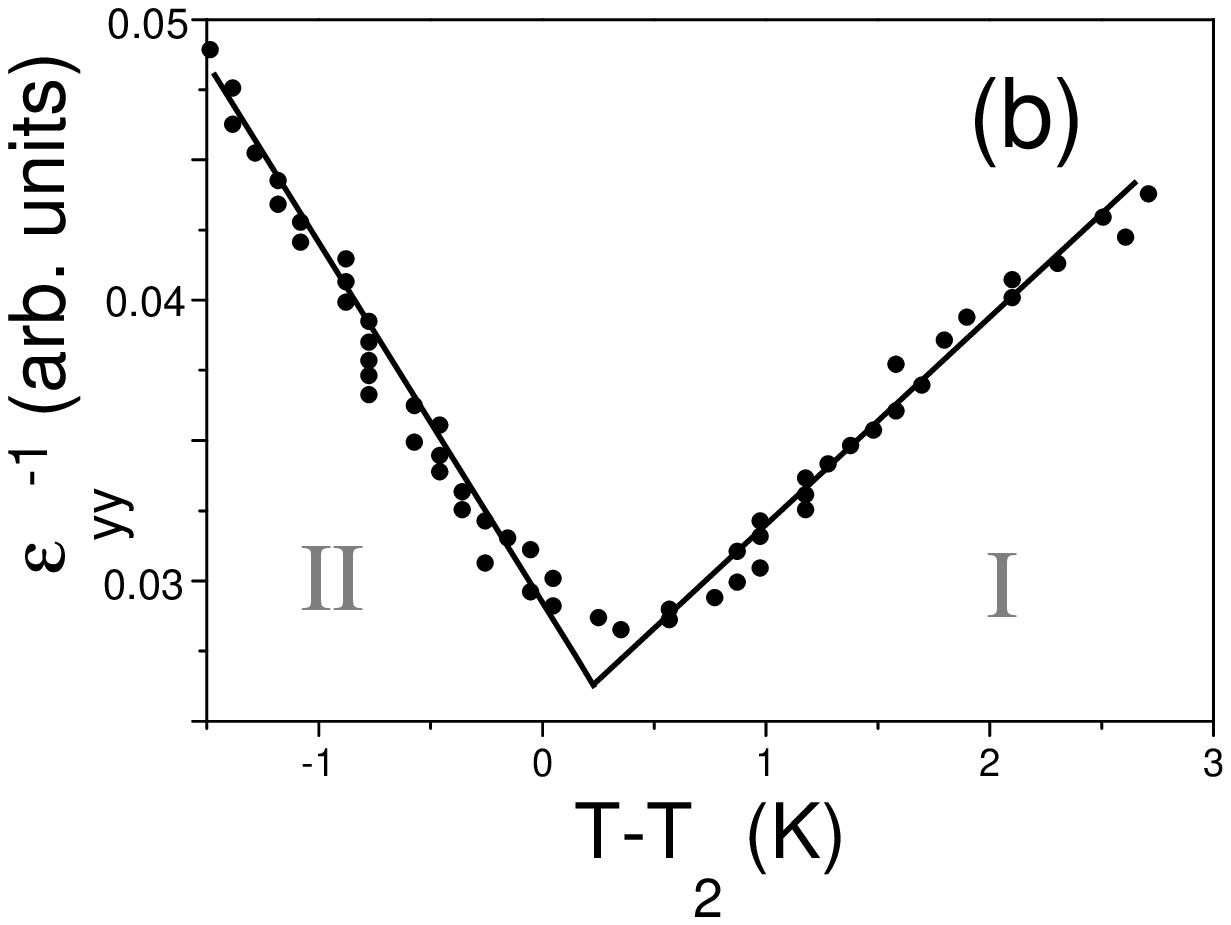}
\includegraphics[scale=0.35]{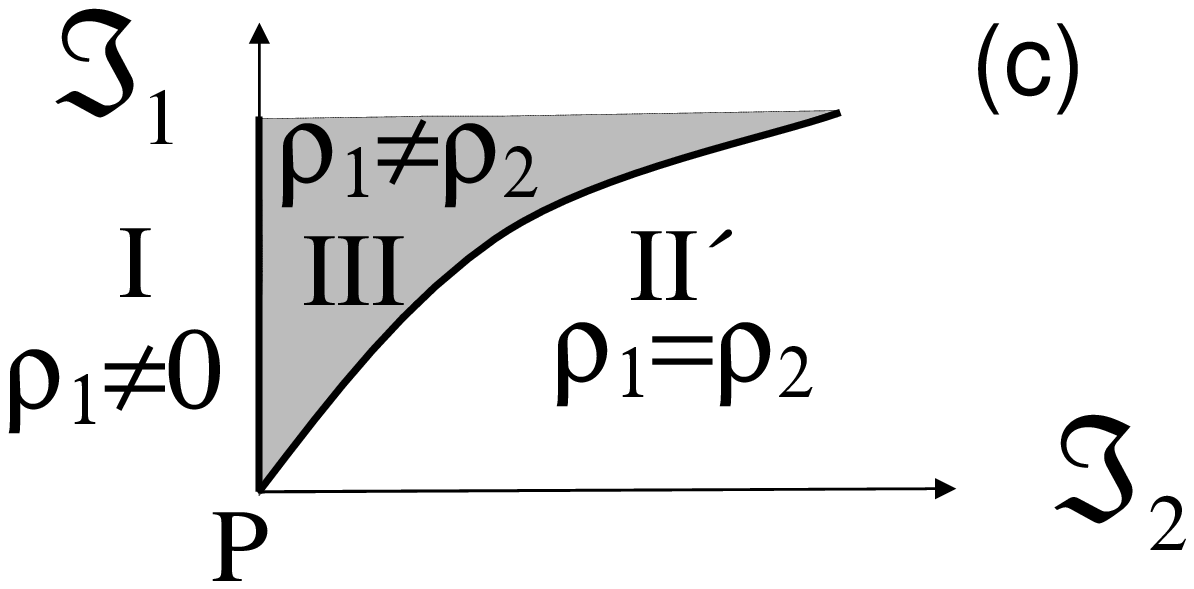}
\includegraphics[scale=0.35]{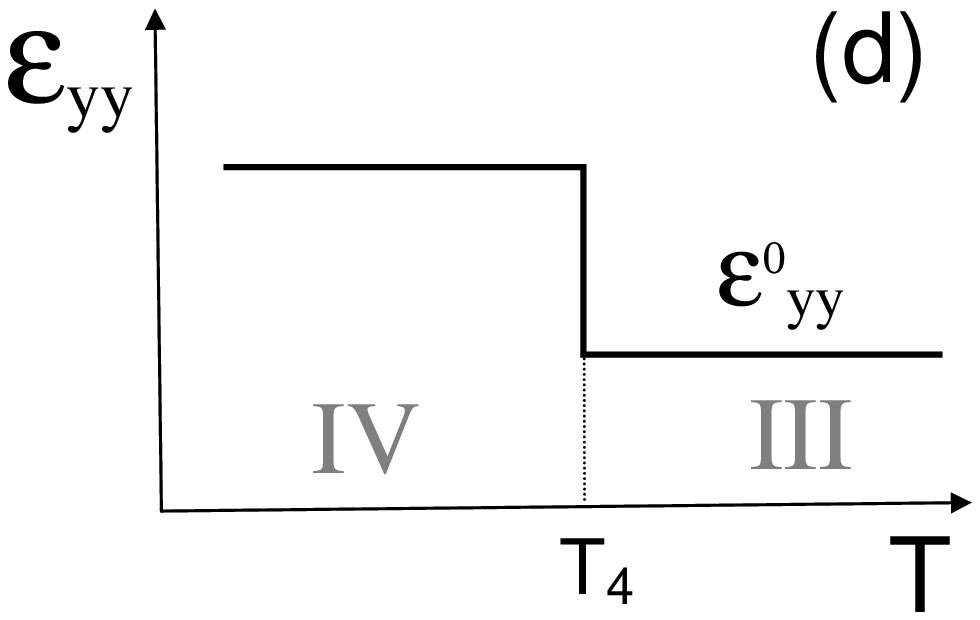}
\caption{(a) Least squares fits for the squared polarization $P_y^2 \propto (T_2-T)$  and (b) the inverse dielectric permittivity $\epsilon_{yy}^{-1} \propto |(T_2-T)|$ reported by Hur et al\cite{Hur1} . (c) Phase diagram associated with $\Xi_1$ in the orbit space ($\mathfrak{I}_1$, $\mathfrak{I}_2$). The structural point-groups are $2_y/m_y$ (phase I), $2_z/m_z$ (phase II) and $\overline{1}$ (phase III). (d) Dielectric permittivity $\epsilon_{yy}(T)$ at the III $\rightarrow$ IV transition.}
\label{fig:3}
\end{center}
\end{figure}

At $T_3=33$ K the wave vector locks into the commensurate value $\vec{k}=(\frac{1}{2},0,\frac{1}{4})$. Table I shows that the symmetry of the $(S_1, S^*_1,S_2,S^*_2)$ order-parameter remains unchanged at the lock-in transition, with the exception of the matrix of the translation $(E|00c)$, allowing formation of the additional (Umklapp-) invariant $\rho_1^4cos4\theta_1+\rho_2^4cos4\theta_2$, which triggers the onset of the commensurate phase.
Therefore, Fig.~1 includes the lock-in commensurate phase III, which involves a fourfold multiplication of the c-lattice parameter, instead of an incommensurate modulation along $z$.
The onset of phase III is reflected by a slight change in the slope of the polarization, with no noticeable anomaly of the dielectric permittivity. It suggests that the magnetic symmetry m'2'm of phase II remains unchanged in phase III.

\section{The III$\rightarrow$IV$\rightarrow$V Transitions}

The wave-vector ${\vec k}=(k_x,0,k_z)\approx (0.48,0,0.32)$ associated with the III$\rightarrow$IV commensurate-incommensurate transition occurring at $T_4$, corresponds to two 4-dimensional IC´s of Pbam1', denoted $\Xi_1$ and $\Xi_2$, whose generators are given in Table I. The four-component order-parameters spanning the two IC´s can be written ($\eta_1=\rho_1e^{i\phi_1}$, $\eta^*_1=\rho_1e^{-i\phi_1}$, $\eta_2=\rho_2e^{i\phi_2}$, $\eta^*_2=\rho_2e^{-i\phi_2})$ for $\Xi_1$  and   ($\varsigma_1=\rho_3e^{i\phi_3}$, $\varsigma^*_1=\rho_3e^{-i\phi_3}$, $\varsigma_2=\rho_4e^{i\phi_4}$, $\varsigma^*_2=\rho_4e^{-i\phi_4}$) for $\Xi_2$. It yields the following independent order-parameter invariants:
($\mathfrak{I}_1=\rho_1^2+\rho_2^2$, $\mathfrak{I}_2=\rho_1^2\rho_2^2$) for $\Xi_1$ , and ($\mathfrak{I}_3=\rho_3^2+\rho_4^2$, $\mathfrak{I}_4=\rho_3^2\rho_4^2$) for $\Xi_2$.\\
Minimization of the free energy associated with $\Xi_1$ yields three possible stable states, shown in the
$(\mathfrak{I}_1,\mathfrak{I}_2)$ phase diagram of Fig.~3(c), which display the \textit{non-polar} structural symmetries $2_y/m_y (\rho_1\neq 0,\rho_2=0), 2_z/m_z (\rho_1=\rho_2)$ and ${\bar 1} (\rho_1\neq \rho_2\neq0)$. The same non-polar symmetries are induced by $\Xi_2$. Therefore, ferroelectric phases IV and V may only result from a \textit{coupling} of the ($\eta_i$) and ($\varsigma_i$) order-parameters associated with $\Xi_1+\Xi_2$, consistent with the observation by Koo et al.\cite{Koo} of a multiple magnetic ordering in phase IV. Taking into account the coupling invariant
$\mathfrak{I}_5=\rho_1^2\rho_3^2\textrm{cos}2\Psi_1+\rho_2^2\rho_4^2\textrm{cos}2\Psi_2$, with $\Psi_1=\Phi_1-\Phi_3$ and $\Psi_2=\Phi_2-\Phi_4$, the free-energy associated with $\Xi_1+\Xi_2$ reads:

\begin{equation}
\label{freenergyx1+xi2}
\Phi_2(\rho_i,\Psi_i)=\sum_{i=1}^5 (\alpha_i\mathfrak{I}_i+\beta_i\mathfrak{I}_i^2)
\end{equation}

Minimization of $\Phi_2$ shows that not less than 15 distinct phases can be stabilized for different equilibrium values of $\rho_i$ and $\Psi_i$. 6 of these phases display a ferroelectric polarization component $P_y$, resulting from the mixed coupling invariant:
$\mathfrak{I}_6=P_y(\rho_1\rho_3\textrm{sin}\Psi_1+\rho_2\rho_4\textrm{sin}\Psi_2)$.
For $\rho_1=\rho_2$, $\rho_3=\rho_4$, $\Psi_1=(2n+1)\frac{\pi}{2}$ or (and) $\Psi_2=n\frac{\pi}{2}$ the phases have the structural symmetry m2m.
For $\rho_1 \neq 0,\rho_3 \neq 0,\rho_2=\rho_4=0$ or $\rho_1=\rho_3=0,\rho_2 \neq 0, \rho_4 \neq 0$ or
$\rho_1 \neq \rho_2$, $\rho_3 \neq \rho_4$ with $\Psi_1$ or $\Psi_2 = (2n+1)\frac{\pi}{2}$ and $\Psi_1$ or $\Psi_2$ arbitrary, or $\Psi_1$ and $\Psi_2$ arbitrary, the structural symmetry is lowered to $2_y$. The magnetic order in the different phases is expressed by the coupling invariants
$\mathfrak{I}_7=M_xM_y(\rho_1\rho_3\textrm{cos}\Psi_1+\rho_2\rho_4\textrm{cos}\Psi_2)$ and  $\mathfrak{I}_8=M_yM_z(\rho_1\rho_3\textrm{cos}\Psi_1-\rho_2\rho_4\textrm{cos}\Psi_2)$.\\
The experimental results reported for the magnetic structure of phase IV of
$\textrm{TbMn}_2\textrm{O}_5$ \cite{Koo} are consistent with a structural symmetry m2m. One can assume, without loss of generality, that the corresponding equilibrium values of the order-parameters in phase IV are $\rho_1=\rho_2$, $\rho_3=\rho_4$, $\Psi_1=(2n+1)\frac{\pi}{2}$, $\Psi_2=n\pi$. Therefore the dielectric contribution to the free-energy at the III$\rightarrow$IV transition is:
$\Phi_2^D=\pm\delta_2\rho_1\rho_3P_y+\frac{P_y^2}{2\epsilon^0_{yy}}$. It yields
\begin{equation}
\label{pol2}
P^e_y=\pm\delta_2\epsilon^0_{yy}\rho_1\rho_3
\end{equation}

Since both order-parameters $\rho_1$ \textit{and} $\rho_3$  contribute  to the symmetry-breaking mechanism at $T_4$, they both vary as $\propto (T_4-T)^{1/2}$ for $T \le T_4$. Therefore $P_yê$ varies linearly as $(T_4-T)$, which expresses a typical improper ferroelectric behaviour for the III$\rightarrow$IV transition. The dielectric permittivity is given by $\epsilon_{yy}=\epsilon^0_{yy}(1-\delta_2\frac{\partial \rho_1 \rho_3}{\partial E_y})$, where $E_y$ is the applied electric field. It yields
($\epsilon_{yy}=\epsilon^0_{yy}$) for $T>T_4$ , and is approximated by
$\epsilon_{yy} \approx \frac{\epsilon^0_yy}{1-\delta_2^2\epsilon^0_{yy}f(\beta_i,\alpha_5)}$
for $T<T_4$, where $f(\beta_i,\alpha_5)$ represents a combination of phenomenological coefficients of $\Phi_2$ with $0<f(\beta_i,\alpha_5)<1$. Accordingly, $\epsilon_{yy}(T)$ undergoes an \textit{upward step} at $T_4$ (Fig.~3 (d)), as observed experimentally \cite{Hur1,Chapon1}.
Note that the change in the order-parameter symmetry imposes a first-order character to the III$\rightarrow$IV transition, consistent with the lattice anomalies observed at 24 K \cite{Chapon1}.\\
The preceding description allows a straightforward explanation of the observed decrease \cite{Hur1} of the equilibrium polarization $P^e_y$ at zero magnetic field which is starting at about 26 K.
Below $T_4$, $P^e_y$ is the sum of two distinct contributions given by Eqs. (\ref{polarization}) and
(\ref{pol2}).
\begin{equation}
\label{two pol contributions}
P^e_y= \pm \epsilon^0_{yy}[\delta_1 \rho_1 (T_2-T)^{1/2} + \delta_2(T_4-T)]
\end{equation}
where $\rho_1 \propto (T_1-T_2)^{1/2}$. Assuming $\delta_1>0$ and $\delta_2<0$, one can verify that
$P^e_y$ decreases for $T \le T_{max} = T_2-\frac{\delta_1^2(T_1-T_2)}{4\delta_2^2}$. This explanation confirms the conjecture by Hur, et al. \cite{Hur1} that the total polarization is composed by positive and negative components, which appear at $T_2$ and $T_4$, respectively. The opposite signs of  $P^e_y$ in Eq.(\ref{two pol contributions}) correspond to the opposite ferroelectric domains disclosed by the preceding authors under opposite electric fields. The strong increase of $P^e_y$ observed below $T_5$ reflects the positive contribution of phase V to the total polarization. The absence of dielectric anomaly at $T_5$ suggests that the m2m symmetry of phase IV remains unchanged in phase V with an eventual change in the respective values of $\Psi_1$ or (and) $\Psi_2$. \\
It remains to understand why the decrease of $P_y$ is enhanced by application of a magnetic field $H_x$, leading to a change in sign of $P_y$ above a threshold field $H^c_x$ \cite{Hur1,Chapon1}. This can be explained by considering the magnetic and magnetoelectric contributions to the free-energy under $H_x$ field in phase IV:
$\Phi^M_2=\mu_0\frac{M_x^2}{2}-H_xM_x$ and $\Phi^{ME}_2= \nu \rho_1\rho_3P_yM_x^2$.
It yields for the field dependent spontaneous polarization in phase IV:
\begin{equation}
\label{pol phase IV}
P^{IV}_y(H_x)=\pm \epsilon^0_{yy}\rho_1\rho_3\Big(\delta_2 + \nu \mu_0^{-1}H_x^2  \Big)
\end{equation}
For $\nu<0$ the application of an $H_x$ field enhances the negative contribution $P^{IV}_y$ to the temperature dependence of the total polarization leading to:
\begin{eqnarray}
\label{temp dep of pol in phase IV}
&&P^e_y(T,H_x)=  \\ 
&&\pm \epsilon^0_{yy}\Big[\delta_1(T_1-T_2)^{1/2}(T_2-T)^{1/2} + (\delta_2 + \nu \mu_0^{-1}H_x^2(T_4-T) \Big] \nonumber
\end{eqnarray}
Accordingly $P^e_y(T,H_x)$ changes sign for the temperature dependent threshold field:
\begin{equation}
\label{threshold field in phase IV}
H^c_x(T)^2=\frac{\delta_1 \mu_0(T_1-T_2)^{1/2}(T_2-T)^{1/2}}{|\delta_2 \mu_0+\nu|(T_4-T)}
\end{equation}
From Eqs. (\ref{temp dep of pol in phase IV}) and (\ref{threshold field in phase IV}) one can verify that with increasing applied field, $P^e_y(T,H_x)$ decreases more sharply and changes sign at higher temperature (Fig.4), as was actually observed by Hur et al. \cite{Hur1}.

\begin{figure}
\begin{center}
\includegraphics[scale=0.55]{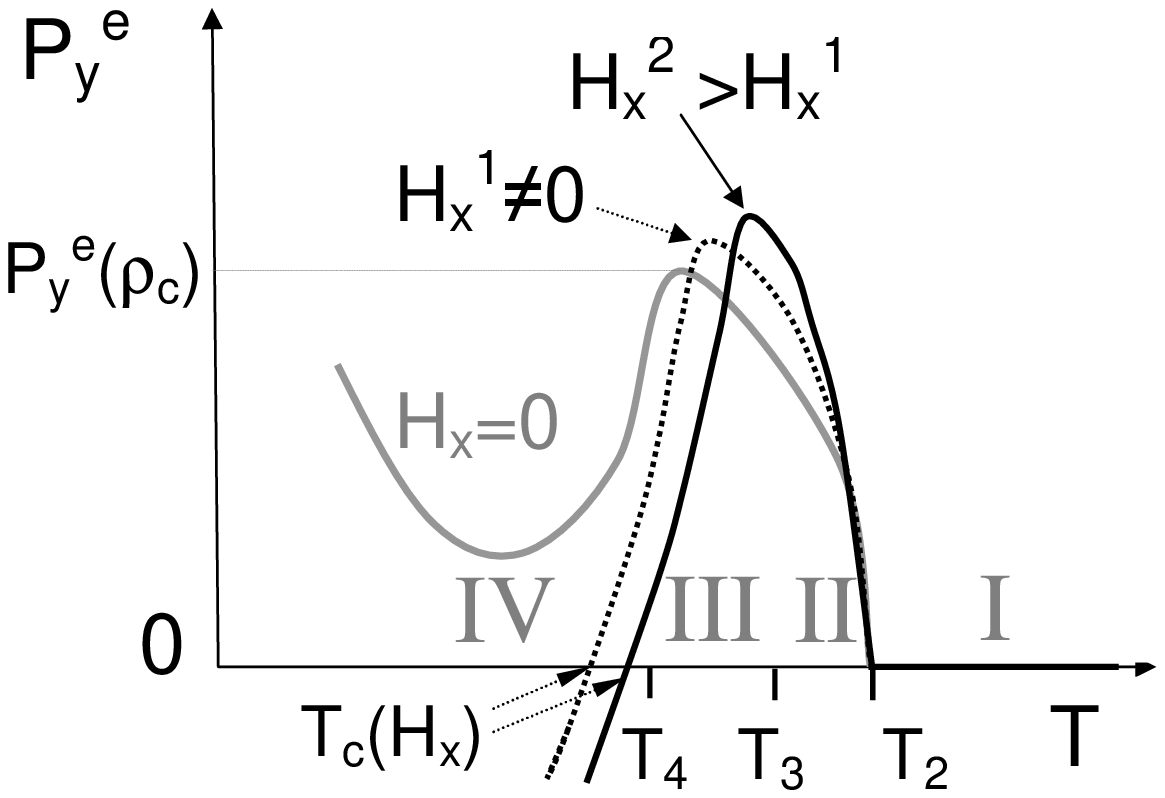}
\caption{Temperature dependence of $P^e_y(T,H_x)$ given by Eq.~(\ref{temp dep of pol in phase IV}), for $\delta_1>0$, $\delta_2<0$ $\nu<0$. With increasing field $P^e_y(T,H_x)$ decreases more sharply and is shifted to higher temperature. The change in sign of $P^e_y(T,H_x)$ occurs at a field-dependent critical temperature $T_c(H_x)$.}
\label{fig:4}
\end{center}
\end{figure}

\section{Summary and Discussion}

  In summary, it has been shown that two distinct symmetry breaking ordering parameters are involved in the sequence of five magnetic phases found in $\textrm{TbMn}_2\textrm{O}_5$  below:\\
1) a single four-component order-parameter is associated with the P$\rightarrow$I$\rightarrow$II$\rightarrow$III transitions.
Two among the components ($S_1,S^*_1$) give rise at $T_1$ to the antiferromagnetic phase I, whereas the two others ($S_2,S^*_2$) are activated at $T_2$ , at the onset of the ferroelectric phase II, ($S_1,S^*_1$) being frozen at the I II transition. It results in a hybrid pseudo-proper ferroelectric behavior for this transition, which displays critical dielectric anomalies typical of proper ferroelectric transitions, although the magnitude of the induced polarization in phase II is of the order found in improper ferroelectrics. At the II$\rightarrow$III transition the translational symmetry along $z$ becomes commensurate, modifying in a minor way the ferroelectric properties of the material. The theoretical phase diagram  showing the location of the phases stabilized in $\textrm{TbMn}_2\textrm{O}_5$, as  well as the other five phases induced by the ($S_i,S^*_i$) order-parameter, has been worked-out, and the magnetic point-groups of the different phases have been given.\\
2) At the commensurate-incommensurate III$\rightarrow$IV transition the ($S_i,S^*_i$) order-parameter splits into two distinct four-component order-parameters ($\eta_i$) and ($\varsigma_i$), which couple for inducing the ferroelectric phases IV and V. The III$\rightarrow$IV transition shows a standard improper ferroelectric behavior. The absence of noticeable anomaly for the dielectric permittivity at the IV$\rightarrow$V transition suggests that the structural symmetry of phase IV remains unchanged in phase V. However, the spontaneous polarization in phase V contributes positively to the observed total polarization  $P^e_y$, whereas phase IV exhibits a negative contribution to $P^e_y$. Opposite signs of the spontaneous electric polarizations in phases IV and V provide a consistent interpretation of the non-monotonous temperature dependence of $P^e_y(T)$ across the II$\rightarrow$III$\rightarrow$IV$\rightarrow$V sequence of induced ferroelectric transitions. Application of an $H_x$ magnetic field modifies the preceding behavior, via the magnetoelectric coupling between $P_y$  and the induced magnetization $M_x$, which contributes negatively to the total polarization, explaining the observed change of sign of $P^e_y(T,H_x)$.\\
A number of previous studies \cite{Radaelli2007,Radaelli2008,Harris2007,Harris2008a,Harris2008b,Kadomtseva,Sushkov}
proposed a theoretical description of the dielectric and magnetoelectric properties of $\textrm{TbMn}_2\textrm{O}_5$. However, none of these studies took fully into account the order-parameter symmetries associated with the different phases. Therefore, the relevant free-energies, expanded to the necessary degrees, and the related coupling terms, could not be disclosed, and the proper phase diagrams could not be derived. As a consequence, a consistent interpretation of the dielectric behavior at zero magnetic field, or under applied $H_x$ field, could not be given explicitly. In contrast to our phenomenological description, based on the symmetry and thermodynamic considerations underlying the Landau theory of magnetic phase transitions \cite{Landau,Dzialo,Toledano1}, which is free from any microscopic model, the previous works, using different group-theoretical procedures, attempted to deduce the magnetoelectric properties of the material from its complex magnetic structures and related magnetic interactions. For example, Radaelli and Chapon limit their group-theoretical analysis to the irreducible corepresentations of the little group \cite{Radaelli2007}. It does not allow determination of the transition free-energy and of the coupling relating the polarization to the magnetic order-parameter, which they deduce from microscopic coupling mechanisms \cite{Radaelli2008}. The complex procedure proposed by Harris \cite{Harris2007} for determining the transition order-parameter from the spin configuration of $\textrm{TbMn}_2\textrm{O}_5$  does not provide the relevant order-parameter symmetry, and is not related organically with the different effective free-energies used by Harris et al \cite{Harris2008a,Harris2008b} for describing the ferroelectric transitions in this material. It leads to oversimplified phase diagrams and to a speculative interpretation of the observed dielectric and magnetoelectric properties. Besides, the critical wave vector assumed by Harris et al. \cite{Harris2008a} for phases I and II of $\textrm{TbMn}_2\textrm{O}_5$ corresponds actually to phases IV and V, which are not described by these authors. The Landau model used by Kadomtseva et al. \cite{Kadomtseva} provides an insight into the exchange and relativistic magnetic contributions to the free-energy and induced polarization involved in $\textrm{RMn}_2\textrm{O}_5$  compounds. However, the  dimensionality of the irreducible representation and order-parameter assumed in their model (2-dimension instead of two coupled 4-dimensional order-parameters required for $\textrm{ErMn}_2\textrm{O}_5$  and $\textrm{YMn}_2\textrm{O}_5$) and the fact that two  successive and distinct order-parameters are needed for describing the full sequence of observed phases, do not allow these authors to describe consistently the observed dielectric properties and magnetoelectric  effects. Along another line, the model proposed by Sushkov et al. \cite{Sushkov} gives an interesting analysis of the underlying magnetic forces explaining the induced dielectric properties in the $\textrm{RMn}_2\textrm{O}_5$ family, but a detailed description of the observed phase sequences and magnetoelectric effects, that would require considering the actual order-parameter symmetries, is missing.\\
Similar sequences of ferroelectric phases are found in other $\textrm{RMn}_2\textrm{O}_5$  compounds \cite{Noda,Chapon2,Bodenthin,Higashiyama,Hur2004,Munoz} where $R=Bi,Y$ or a rare-earth heavier than $Nd$. In these compounds, with the exception of $\textrm{BiMn}_2\textrm{O}_5$, the first ferroelectric phase does not appear directly below the paramagnetic phase, but below an intermediate non-polar antiferromagnetic phase. Therefore the induced electric polarization results from the coupling of two distinct magnetic order-parameters, one of which having been already activated in the intermediate phase. As a consequence a pseudo-proper coupling is created, which gives rise to the typical critical behavior of a proper ferroelectric transition. In $\textrm{TbMn}_2\textrm{O}_5$  the first ferroelectric transition corresponds to ${\vec k}=(\frac{1}{2},0,k_z)$  and to a \textit{single} 4-component order-parameter, the pseudo-proper coupling occurring between \textit{distinct components} of the same order-parameter. A different situation is found in the other $\textrm{RMn}_2\textrm{O}_5$  compounds, in which the first transitions correspond to ${\vec k}=(k_x,0,k_z)$ \cite{Noda,Chapon2,Bodenthin,Higashiyama,Hur2004,Munoz}, i.e. the pseudo-proper coupling occurs between \textit{two distinct 4-component order-parameters} having the symmetries of the ($\eta_i$) and ($\varsigma_i$) order-parameters, which are associated in the present work to the lower temperature transition sequence  of $\textrm{TbMn}_2\textrm{O}_5$.
The order-parameters involved in the $\textrm{RMn}_2\textrm{O}_5$ family correspond in most cases to the symmetries disclosed in the present work for $R=Tb$, i.e. to ${\vec k}=(\frac{1}{2},0,k_z)$, $(k_x,0,k_z)$  and $(\frac{1}{2},0,\frac{1}{4})$.
Two exceptions are presently known, which are:\\
 1) The lower temperature phase of $\textrm{DyMn}_2\textrm{O}_5$ \cite{Hur2004}  induced by bidimensional order-parameters corresponding to the wave-vector $(\frac{1}{2},0,0)$, and \\
 2) The single ferroelectric phase of  $\textrm{BiMn}_2\textrm{O}_5$ \cite{Munoz}induced by bidimensional IC's at ${\vec k}=(\frac{1}{2},0,\frac{1}{2})$. \\
 Accordingly, despite the apparent variety of behaviors found for the dielectric properties and field effects a unifying theoretical framework can be proposed for the $\textrm{RMn}_2\textrm{O}_5$  manganites, which can be deduced from the description given in the present work, by interchanging the order-parameters in the observed transition sequences.
 
 \section{Conclusions}
 
 In conclusion, the order-parameter symmetries associated with the magnetostructural transitions observed in  $\textrm{TbMn}_2\textrm{O}_5$ clarify the nature of the ferroelectric phases and permit a consistent description of the magnetoelectric effects observed in this material. In a more general way, our phenomenological approach illustrates the necessity of taking into account the actual order-parameter symmetries and phase diagrams associated with the phase sequences reported in multiferroic compounds. It can be used for analyzing the complex microscopic mechanisms and interactions involved in magnetostructural transitions, which have not been discussed in the present work.\\

\acknowledgements
Support by the Austrian FWF (P19284-N20) and by the University of Vienna within the IC Experimental Materials Science ("Bulk Nanostructured Materials") is gratefully acknowledged.

\end{document}